\documentclass[twocolumn,prd,nofootinbib,showpacs,floatfix,superscriptaddress]{revtex4}
\usepackage{epsfig}
\usepackage[dvips]{color}

\def\alt{\raise0.3ex\hbox{$\;<$\kern-0.75em\raise-1.1ex\hbox{$\sim\;$}}}
\def\agt{\raise0.3ex\hbox{$\;>$\kern-0.75em\raise-1.1ex\hbox{$\sim\;$}}}

\definecolor{Black}{named}{Black}
\definecolor{Red}{named}{Red}

\newcommand{\bw}{\begin{widetext}}
\newcommand{\ew}{\end{widetext}}
\begin{document}
\def\d{{\rm d}}
\def\ap{\approx}
\def\theta{\vartheta}
\def\Ecut{E_{\rm cut}}
\def\C{{\mathcal C}}

\title{Clustering properties of ultrahigh energy cosmic rays
and the search for their astrophysical sources}

\author{A.~Cuoco}
\author{S.~Hannestad}
\author{T.~Haugb{\o}lle}
\address{Department of Physics and Astronomy, University of
Aarhus, Ny Munkegade, DK--8000 Aarhus, Denmark}

\author{M.~Kachelrie\ss}
\address{Institutt for fysikk, NTNU, N--7491 Trondheim,
Norway}

\author{P.~D.~Serpico}
\address{Center for Particle Astrophysics, Fermi National
Accelerator Laboratory, Batavia, IL 60510-0500, USA}

\date{\today}

\begin{abstract}
The arrival directions of ultrahigh energy cosmic rays (UHECRs)
may show anisotropies on all scales, from just above the
experimental angular resolution up to medium scales and dipole
anisotropies. We find that a global comparison of the two-point
auto-correlation function of the data with the one of catalogues
of potential sources is a powerful diagnostic tool. In particular,
this method is far less sensitive to unknown deflections in
magnetic fields than cross-correlation studies while keeping a
strong discrimination power among source candidates. We illustrate
these advantages by considering ordinary galaxies, gamma ray
bursts and active galactic nuclei as possible sources. Already the
sparse publicly available data suggest that the sources of UHECRs
may be a strongly clustered sub-sample of galaxies or of active
galactic nuclei. We present forecasts for various cases of source
distributions which can be checked soon by the Pierre Auger
Observatory.
\end{abstract}

\pacs{98.70.Sa,   
      98.54.Cm    
\hfill FERMILAB-PUB-07-467-A}
\maketitle

\section{Introduction}
The identification of the sources of ultrahigh energy cosmic rays
(UHECRs) and, more generally, the question if astronomy with charged
particles is possible are two important unresolved problems of
astroparticle physics. The answer to the latter question depends
both on the magnitude of deflections in magnetic fields (which in
turn depends also on the chemical composition of UHECR primaries)
and on the number density and the luminosity of UHECRs sources.
Consensus has not yet emerged on the origin and the amplification
mechanisms of primordial magnetic fields, nor on the present
magnitude and structure of extragalactic magnetic fields outside of
galaxy cluster cores~\cite{ems,dgst}. Uncertainties from modeling strong
interactions prevent a clean determination of the fraction of heavy
nuclei in UHECRs above $E\agt 10^{18}\,$eV~\cite{chemie}. Thus
theoretical predictions about the chances of charged particle
astronomy differ drastically, and the answer has to come from
experiment.

There are various  pieces of evidence in the available
experimental data. The AGASA data contain several small-scale
clusters, i.e.\ clusters of events within its experimental angular
resolution~\cite{ssc}. This result triggered a series of works
studying the auto-correlation of UHECR data at small angular
scales~\cite{auto,finite} or correlating UHECRs with potential
astrophysical sources~\cite{corr}. For instance, the best-fit
value for the density $n_s$ of UHECR sources found in
Ref.~\cite{finite} is $(1-3)\times 10^{-5}/$Mpc$^3$, while the
$2\:\sigma$ confidence region ranges from $2\times
10^{-6}/$Mpc$^3$ to $\sim 10^{-2}/$Mpc$^3$, i.e.\ up to the
density of galaxies. The large statistical error of this estimate
comes mainly from the small number of doublets with less than
3--5~degrees separation, while deflections in magnetic fields of
more than a few degrees would result in a systematic
overestimation of $n_s$. Correlation analyses of the small UHECR
data set have their own problems: In order to avoid a too large
number of potential sources per angular search bin, one has to
choose either a very high energy cut or a very specific test
sample, e.g.\ a small subset of all Active Galactic Nuclei (AGN).
Although some studies found significant correlations, in
particular with BL Lacs, these results have remained
controversial.

A second piece of evidence are anisotropies on medium scales. The
authors of Ref.~\cite{msc} analyzed the available data set of CR
arrival directions from the  HiRes stereo, AGASA, Yakutsk and
SUGAR experiments with energies $E\geq 4\times 10^{19}\:$eV in the
HiRes energy scale. They found evidence at $\sim 3\:\sigma$~C.L.
for anisotropies on the scales of 10--35$\:$degrees, with a clear
minimum of the chance probability around 20--30$\:$degrees. This
result is consistent with the theoretical expectations for
anisotropies  associated with large-scale structures (LSS) from
Ref.~\cite{nI}. Further studies showed that the correlations are
best explained if UHECR sources are either over-biased with
respect to normal galaxies and/or if the cosmic ray horizon is
smaller than expected for rectilinearly propagating
protons~\cite{nII}.

Intriguingly, similar findings seem to emerge from an analysis of
the preliminary data from the Pierre Auger Observatory (PAO). For
64 events with $E>4\times 10^{19}\:$eV the data presented in
Ref.~\cite{ICRC} show a surplus of clustering in the broad range
from 7 to 30 degrees. The distribution has its minimum at
7~degrees with a second, broad minimum between 19--24~degrees and
is quite similar to the distribution with 57~events in
Ref.~\cite{msc}. Remarkably, the PAO data contain also ``8
doublets separated by less than 7 degrees in the 19 highest energy
events'', i.e. with $E>5.75\times 10^{19}\:$eV~\cite{ICRC}.

Finally, the last piece of evidence comes from the UHECRs energy
spectrum. In particular, the long controversy  about the
continuation of the spectrum \cite{Takeda:1998ps} beyond the
Greisen-Zatsepin-Kuzmin (GZK)
cutoff \cite{Greisen:1966jv,Zatsepin:1966jv} seems to be finally
solved by the latest data from HiRes \cite{Abbasi:2007sv} and the
PAO \cite{Yamamoto:2007xj} that both detect with an high confidence
level ($>5 \sigma$) a prominent steepening in the spectrum
compatible with the GZK attenuation. Complemented by the new PAO
stringent limit on the fraction of UHECRs photon primaries
\cite{Semikoz:2007wj} the data are clearly pointing toward a
``standard" scenario in which the bulk of UHECRs sources have an
astrophysical origin in the nearby universe, with more exotic
top-down scenarios playing at most a sub-dominant role. This
evidence makes timely a detailed study of possible UHECRs
astrophysical sources of the kind addressed in the following.

The main aim of the present work is to compare the auto-correlation
function of potential UHECR sources with these early results of the
PAO on UHECR arrival directions, and to provide forecasts of the
clustering expected for different classes of sources which can be
checked shortly by the PAO. We shall also comment on how the
clustering features of the public available world data-set compare
with expectations. The timeliness of this analysis is due to the
fact that, while increasing evidence is accumulating in favor of an
astrophysical origin of UHECRs, it is still unclear how accurately
one can identify the sources of UHECRs and what are the best tools
to do so. Here we advocate the importance of a global comparison,
i.e.\ a comparison on all angular scales, of the observed
auto-correlation function of arrival directions with the expectations
for different source and primary scenarios. At first glance,
cross-correlation tests with source catalogues might appear as
the ideal tool to identify the UHECR sources, but if used alone they
could be insufficient or misleading. First, the angular resolution of
${\mathcal O}$(1$^\circ$) of UHECR observatories is poor for astronomical
standards. Additionally, UHECRs are plagued by non-negligible
magnetic field deflections, except maybe at the highest energies
observed and for proton primaries. Non-spurious signals in a
cross-correlation analysis can only be expected if the {\it overall\/}
magnetic deflection of UHECRs is below the size of the angular bin
used. Small-scale auto-correlation studies are less sensitive to
magnetic fields since only the relative deflections between pairs of
events enter. Yet, since the overall spreading induced by the magnetic fields
is unknown, an auto-correlation analysis limited to the first
angular bin (whose size is chosen a priori, e.g. motivated by the
angular resolution of the observatory) may be unsuccessful and/or have
an ambiguous interpretation. By definition, however, if UHECR astronomy
is possible at all---at least in a statistical sense---sufficiently
large angular scales in the auto-correlation function should reflect
the analogous properties of the sources. As a more ambitious goal, a
global analysis also offers the possibility to infer
the average size of deflections at the chosen energy scale, and thus
to perform a kind of magnetic field reconstruction.

For this line of reasoning to be effective, one has to show first
that the auto-correlation functions of different potential UHECR
sources differ significantly and thus might be used to identify the
UHECR sources. This task is addressed in Sec.~\ref{astrocat}, where
we discuss several astronomical catalogues. For the study of UHECR
anisotropies, catalogues should cover the largest possible fraction
of the sky, ideally complete in distance up to redshift $z\sim
0.1$. We shall see that this is rarely the case at present, which
forces us to limit most of our quantitative comparisons to the sample of
UHECRs with the highest energies: As a rule of thumb, the higher the
UHECR energy, the smaller the energy-loss horizon (and the magnetic
deflections), and the more reliable the catalogue. Turning this into
quantitative statements requires however some assumptions on the
nature of primary particle and the absolute energy scale of experiments.
In Sec.~\ref{compdata}, we perform a comparison with the
PAO results under the hypothesis of proton primaries and using two
different assumptions on the energy scale. There, we also comment on the
the world data-set of available data from the HiRes, AGASA, Yakutsk,
and SUGAR experiments. Finally, we discuss our results and conclude in
Sec.~\ref{conclusion}.

\begin{figure}[!t]
\begin{center}
\begin{tabular}{c}
\epsfig{file=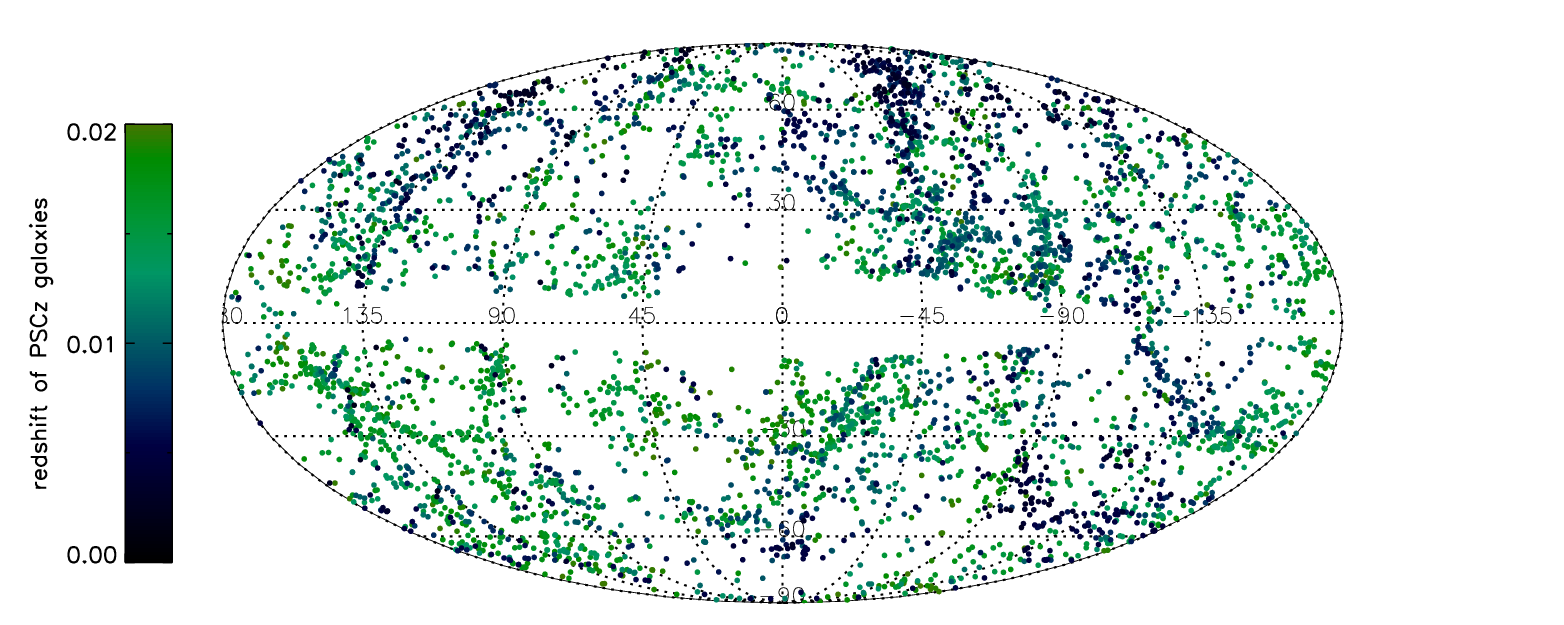,width=10cm} \\
\epsfig{file=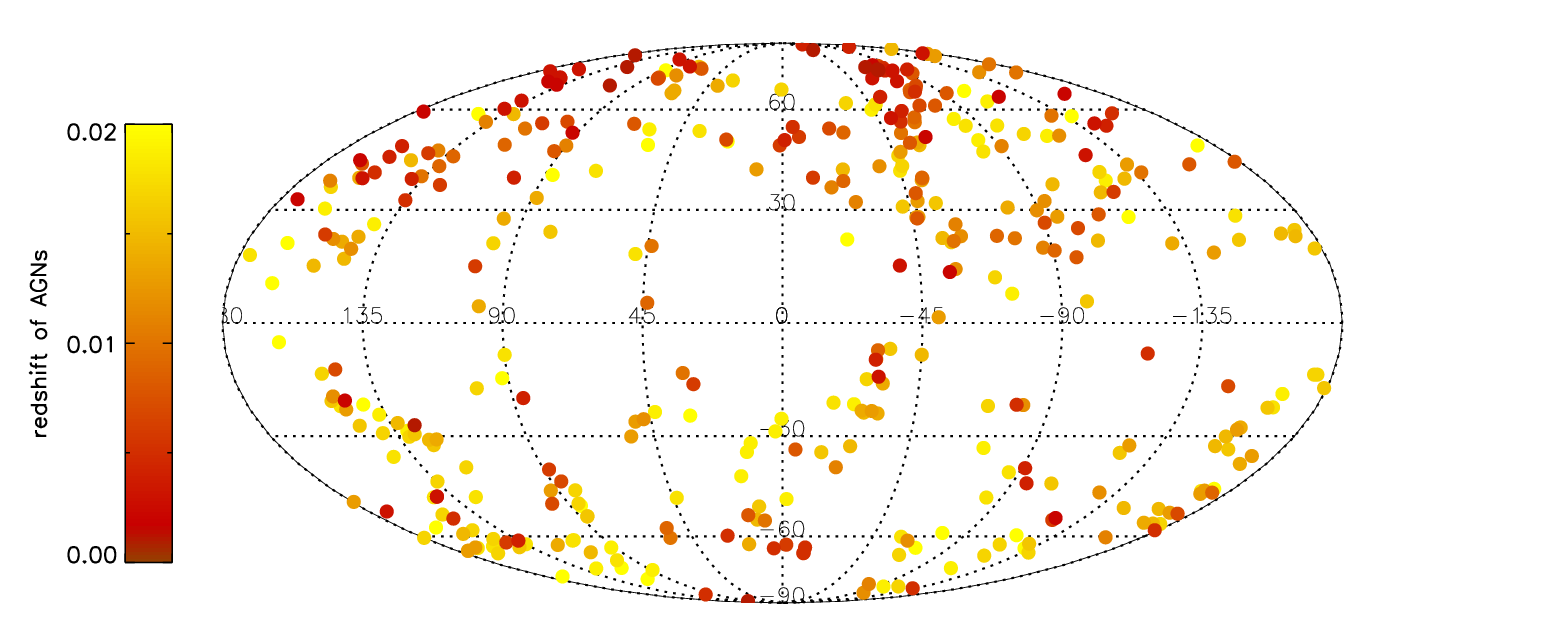,width=10cm}
\end{tabular}
\end{center}
\caption{Top panel: PSCz galaxies within $z=0.02$ color coded in
black to green according to increasing redshift. Bottom panel:
AGNs within $z=0.02$ color coded in red to yellow according to
increasing redshift. Both panels are in Galactic Coordinates.
\label{fig:1}}
\end{figure}

\section{Galaxy and AGN correlation functions}\label{astrocat}
\subsection{Astronomical catalogues}

Among the astrophysical objects most often proposed as UHECR
sources are AGNs in general, specific subclasses like Blazars,
Radio or Seyfert galaxies, Gamma Ray Bursts (GRB) or young neutron
stars (for a review see e.g.~\cite{Torres:2004hk}).  All these
sources follow the LSS of matter, although with different and
scale-dependent biases. In order to understand how large this bias
is for different source classes, we examine now the clustering
properties of normal galaxies (which may host candidates like
neutron stars) and AGNs in the nearby universe.  We use the PSCz
catalogue~\cite{Saunders:2000af} as a sample of the galaxy
distribution and the 12th edition of the V{\'e}ron-Cetty \&
V{\'e}ron (VCV) catalogue~\cite{VC12} for the AGNs. We also study
the clustering properties of several sub-samples, imposing cuts in
absolute magnitude for the galaxy catalogue and subdividing the
AGN catalogue into Seyfert galaxies of type 1 ({\sf S1}), type 2
({\sf S2}) and LINERs ({\sf S3}), according to the classification
reported in VCV catalogue itself. In Fig.~\ref{fig:1} the various
kinds of AGNs and the PSCz galaxies within $z=0.02$ are shown in
Galactic Coordinates, color coded according to their redshift. The
empty region along the Galactic Plane is the so-called avoidance
region due the presence of the Milky Way and does not reflect an
intrinsic lack of objects.

\begin{figure}[t]
\begin{center}
\begin{tabular}{c}
\epsfig{file=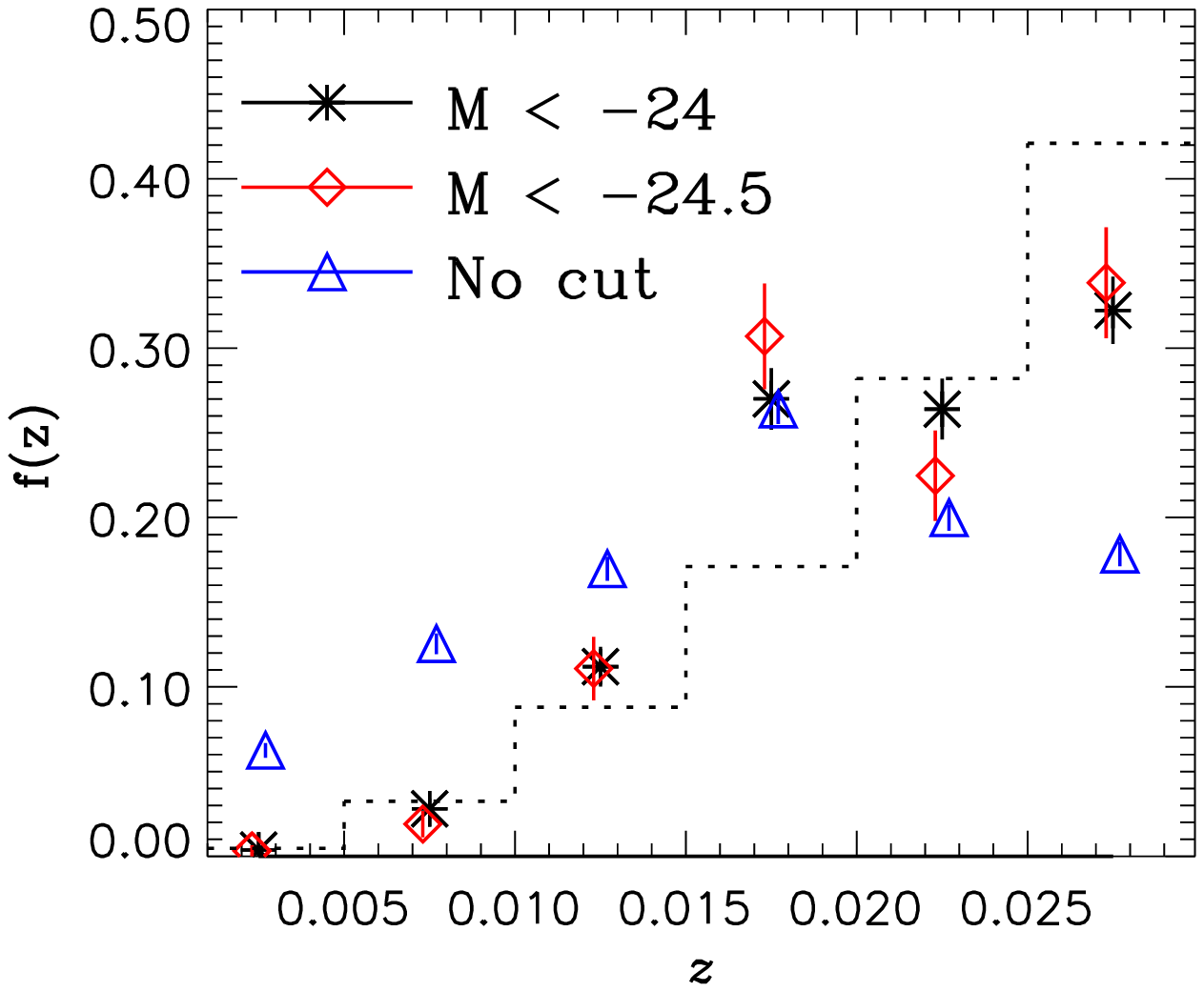,width=7cm}\\
\epsfig{file=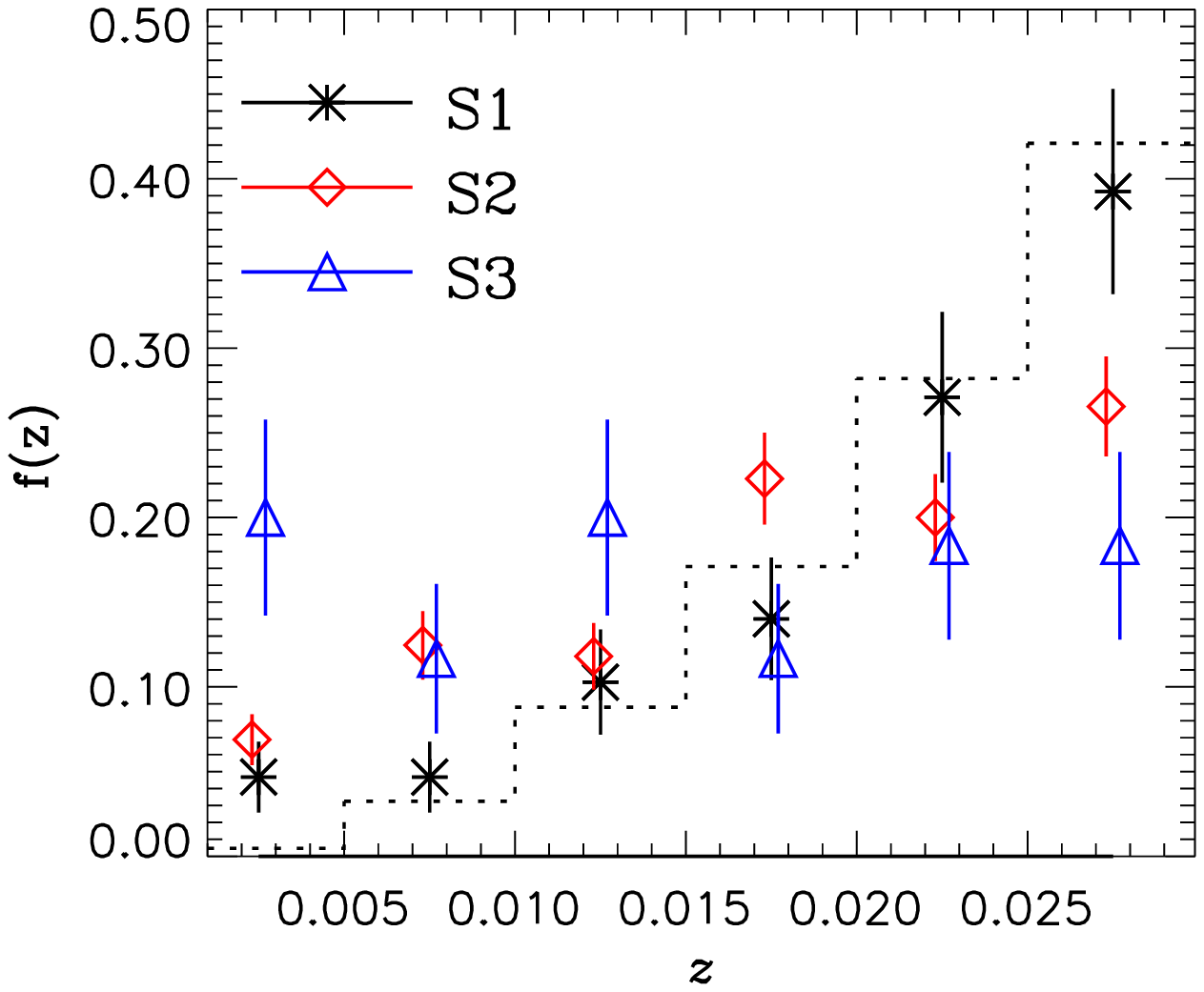,width=7cm}
\end{tabular}
\end{center}
\vspace{-1pc} \caption{Top Panel: The fraction $f(z)$ of galaxies
from the PSCz catalogue is shown for various luminosity cuts $M <
-24, -24.5$ and with no cut in redshift bins of width 0.005.
Bottom Panel: The fraction $f(z)$ of {\sf S1}, {\sf S2}, and {\sf
S3} AGNs. The dotted histogram shows the expected behavior of a
volume-complete catalogue, i.e. with differential fraction
$\propto z^2$. The error bars are the Poisson fluctuations from
the number count. }\label{fig:2}
\end{figure}

The details of the PSCz catalogue, in particular a description of
the mask and of the selection function, are summarized in
Ref.~\cite{Saunders:2000af}. The B-band magnitudes reported in the
catalogue are, however,  biased and show systematic offsets within
different regions of the sky where the galaxy magnitudes have been
taken from different catalogues with different calibrations. To
overcome this problem, we match sources in the PSCz catalogue with
sources in the 2MASS extended source
catalogue~\cite{Jarrett:2000me}, to get accurate magnitudes in the
infrared ($2.15 \, \mu$m) K-band. This is done by requiring that a
PSCz galaxy is inside the 20 mag arcsec$^{-2}$ isophote in the
K-band of the matching 2MASS galaxy. We find that $\sim80$\% of
the galaxies in the PSCz catalogue have a counterpart in the 2MASS
XSC and discard the others. We then construct various sub-samples
of the catalogue performing cuts in absolute magnitude using the
distance modulus relation $M = m-5 \log d_{L,{\rm Mpc}}-25 \sim
m-43.16+5 \log h -5 \log z$ (where $h$ is the reduced Hubble
parameter and  the redshift dependent K-correction, negligible for
$z<0.03-0.04$, has not been considered). For these sub-samples we
also empirically build new selection functions using a smooth
weight function chosen in order to reproduce the redshift
distribution of the sub-sample. In the top panel of
Fig.~\ref{fig:2} we show the fraction $f(z)$ of galaxies from the
PSCz catalogue for the sub-samples obtained with luminosity cuts
$M < -24, -24.5$ in redshift bins of width 0.005. In the same
panel we also show the complete galaxy sample with no cut imposed.
The PSCz catalogue is flux complete and from the figure it can be
seen that the brightest sub-samples are essentially also volume
complete out to $z \sim 0.02$. In contrast, the full sample shows
prominent signatures of volume-incompleteness already at very low
redshift.

\begin{figure*}[!th]
\begin{center}
\begin{tabular}{cc}
\hspace{-2pc} \epsfig{file=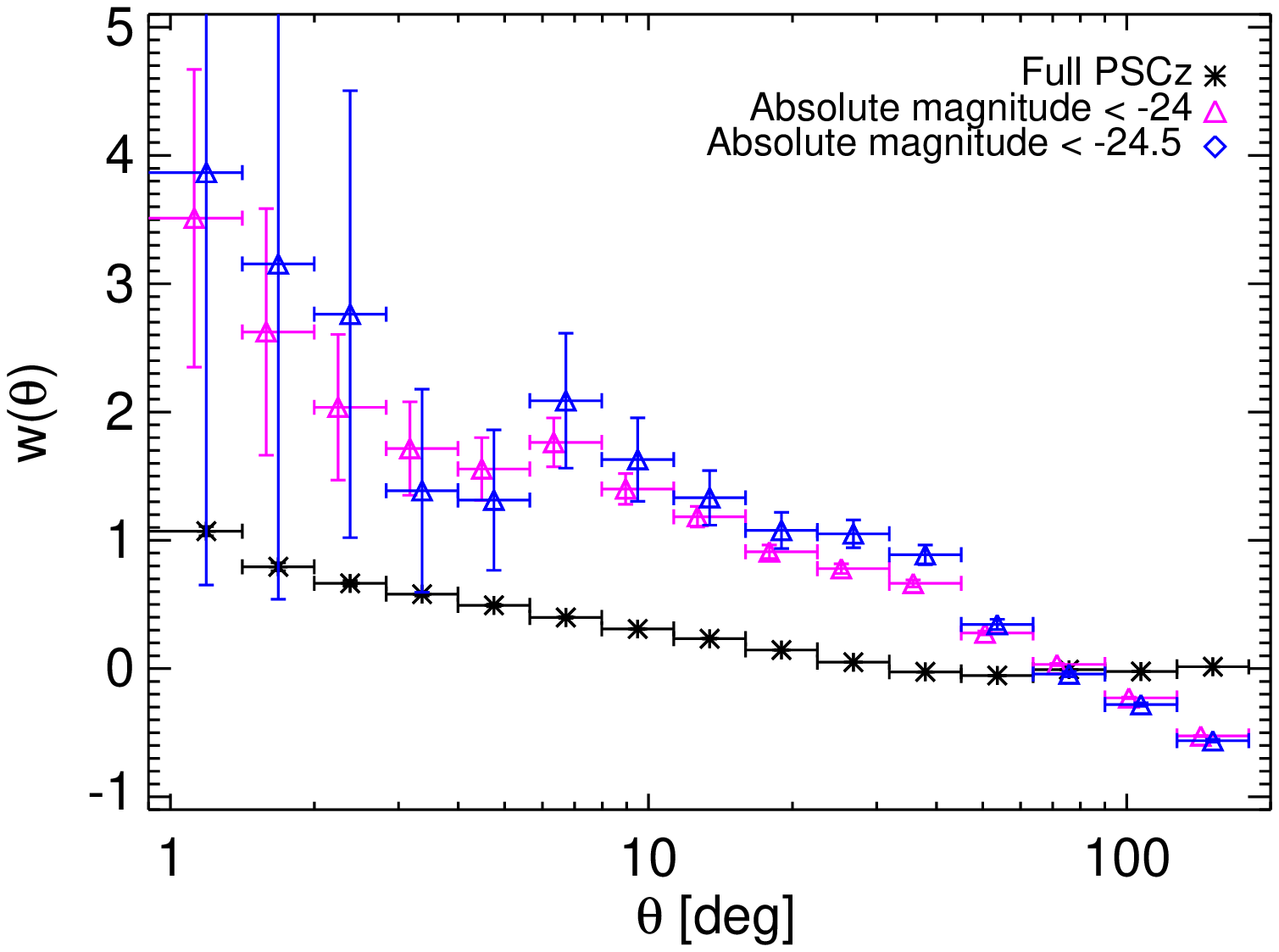,width=9.5cm} \hspace{-2pc} &
\epsfig{file=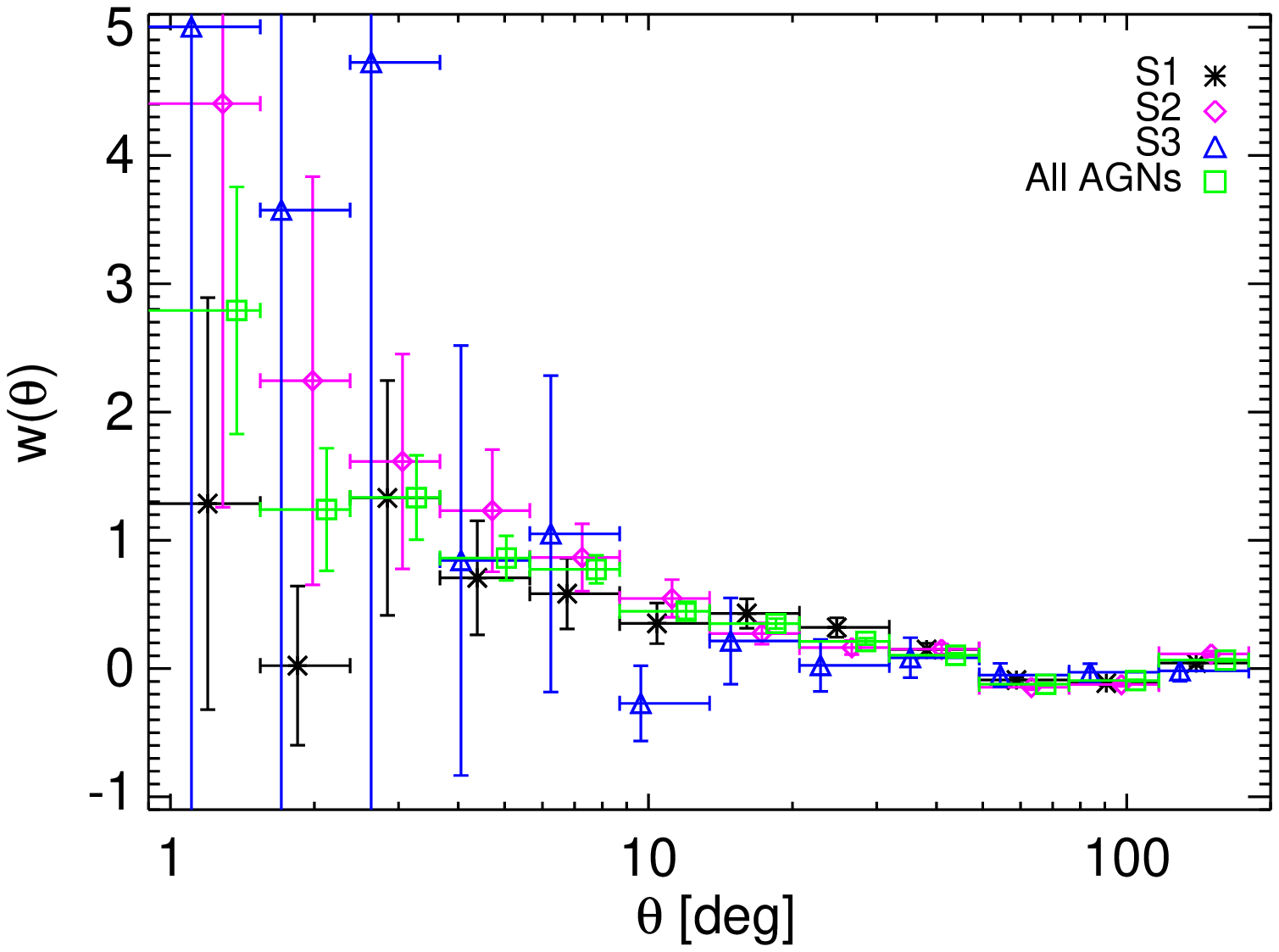,width=9.5cm}
\end{tabular}
\end{center}
\caption{The auto-correlation function $w(\theta)$ for galaxies and
for differents cuts in magnitude (left) and AGNs and different
subsamples (right) as function of  $\theta$ with $1\,\sigma$
error bars.\label{fig:3}}
\end{figure*}

Differently from the PSCz catalogue, the VCV catalogue is a
compilation of observations and is known to suffer increasing
incompleteness with increasing redshift. We assume that at least
in the very nearby universe (i.e. $z \lesssim 0.02$) the catalogue
can be considered fairly complete and we build selection functions
analogously to the PSCz sub-samples described above. It will later
be seen that our results are quite robust against this assumption.
We find that: (i)  within $z=0.02$ the AGN sample follow closely
the matter distribution, as expected (see Fig.\ref{fig:1}); (ii)
although the overall catalogue is fairly complete within this
distance,  there are some subsets of AGNs which suffer from
significant incompleteness and therefore selection bias. In the
bottom panel of Fig.~\ref{fig:2} we show the fraction of {\sf S1},
{\sf S2}, and {\sf S3} galaxies from the VCV catalogue in redshift
bins out to $z=0.03$, together with the Poisson fluctuation due to
the finiteness of the sample (the sample of active galaxies within
$z=0.02$ includes $\sim 500$ AGNs of which $\sim 150$ {\sf S1},
$\sim 200$ {\sf S2}, and  $\sim 80$ {\sf S3}). The {\sf S1} and,
to a minor extent, the {\sf S2} galaxies show a behavior which is
reasonably close to the expected $z^2$ increase of a truly
complete sample. Since they constitute by far the largest fraction
of AGNs in the VCV catalogue, the catalogue as a whole can be
regarded as at least reasonably complete out to $z\sim 0.02$. We
note however that the {\sf S3} sub-sample seems the most affected
by incompleteness effects, a point which is further discussed
below. It is difficult to estimate the number density of AGNs in
the VCV catalogue given that, by construction,  the astrophysical
sources are included without any specific selection rule. As a
rough estimate, assuming volume-completeness with 500 AGNs within
$z=0.02$ gives $n_s\sim 5\times10^{-4}$Mpc$^{-3} h^3$. The same
density corresponds also to the galaxies brighter than $M_{\rm
cut}=-24$, while 50 uniformly distributed sources, always within
the same volume, would give $n_s\sim 5\times10^{-5}$Mpc$^{-3}
h^3$. For ordinary galaxies the number density depends strongly on
the assumed $M_{\rm cut}$ and typically ranges from
$10^{-3}$Mpc$^{-3} h^3$ to values as high as $10^{-1}$Mpc$^{-3}
h^3$ for  Milky Way-like galaxies. In any case, the value of the
number density of the UHECR sources depends on the horizon
containing the sources and thus on issues like the nature of the
primaries and the absolute energy scale. Thus, in the following we
will focus the attention mainly on the absolute number of sources
(above the assumed CR energy threshold) and to their
bias/overdensity with respect the distribution of matter, as
principal observables.

BL Lacs AGNs, also popular UHECR source candidates, are very rare
in our GZK neighborhoods. The nearest confirmed BL Lacs in the VCV
catalogue is the object \textsf{RXS} J05055+0416 at $z=0.027$.
Including also possible BL Lacs only 6 objects are found in the
VCV within $z=0.03$.  If indeed such a small number of sources
would be responsible for the UHECRs even more peculiar clustering
signatures should be expected as a large number of triplets or
even quadruplets, as long as deflections by extragalactic magnetic
fields are not too large.  At the same time, a cross correlation
between these objects and the UHECR multiplets should become
evident. This possibility could then be easily confirmed/disproved
looking at this kind of signatures and we will not discuss it
further in the following.

Finally, we briefly discuss the case of GRBs as UHECR sources. The
observed rate of GRB is $R_{\rm obs}\ap 0.5\times
10^{-9}/($Mpc$^3$yr) according to Ref.~\cite{Schmidt:2001pz}.
However, deflections in the extragalactic magnetic fields (EGMF)
lead to time delays $\tau$ that in turn increase the effective
density of GRBs as $n_s=\tau R_{\rm obs}$. The clustering
properties of GRBs are in general quite different from those of
AGNs and massive galaxies. Long duration GRBs which make up about
2/3 of all GRBs are associated with supernova events in extremely
massive stars and therefore their distribution essentially follows
the star formation rate. Star forming galaxies are mainly spirals
and irregulars which are less clustered than average galaxies in
the PSCz catalogue. The remaining 1/3 of the GRBs which are most
likely the result of binary collisions have a distribution which
is close to the one of SN~Ia's, but are not considered very likely
sites for the UHECR acceleration. Thus GRBs cluster less than
average PSCz galaxies and, in the following considerations, we
shall use randomly distributed sources as a rough template for
their clustering properties.

\subsection{Correlation functions}

An important point to prove for the following arguments is that
different astrophysical catalogues of candidate UHECR sources have
sufficiently different clustering properties. To that purpose, we
calculate in this section  the auto-correlation function
$w(\theta)$ of the various samples. In the past, a commonly employed estimator
for the auto-correlation has been the intuitive $DD/RR-1$. This
estimator is however sub-optimal especially for the estimation of
variance \cite{Landy:1993yu}, while an optimal estimator is given
by~\cite{Blake:2006qu,Landy:1993yu}:
\begin{equation}\label{corrfunc}
    w(\theta)=\left\langle\frac{DD-2DR+RR}{RR}\right\rangle \,,
\end{equation}
where $D$ denotes the data-set and $R$ a randomly generated
data-set with the same bias characteristics as the data (same
mask, same selection function, same exposure, etc.), while the
quantities $DD$, $DR$, $RR$ are the normalized pair counts in each
angular bin around $\theta$. The brackets indicate that the final
$w(\theta)$ is an ensemble average over many random realizations.
Note that for data consistent with a random distribution
$w(\theta)$ is zero within the errors. The resulting
auto-correlation functions $w(\theta)$ are shown in the left panel
of Fig.~\ref{fig:3} for galaxies and in the right panel for AGNs,
without weights for the sources (i.e.\ without selection effects
and attenuation) and using the same masking for all sets in order
to have an unbiased comparison. The errors in each bin can be
estimated as \cite{Landy:1993yu}
\begin{equation}\label{corrfuncerrs}
\textrm{rms}[w(\theta)]=[1 + w(\theta)] \left[\frac{1}{n(n-1)/2\,
\left\langle RR\right\rangle}\right]^{1/2},
\end{equation}
where $n$ is the number of points in the data set $D$, and hence
$n(n-1)/2$ the total number of unique pairs.

Both samples show a strong auto-correlation at small scales,
although the clustering of normal galaxies is quite less
pronounced ($w_{\rm AGN}(1^\circ)/w_{\rm gal}(1^\circ)\simeq 3$).
We will see that in the relation to the small scale clustering
seen by the PAO, this difference already tightly constrains the
possible contribution of normal galaxies as sources of the highest
energy CRs. The situation changes when bright sub-samples of the
PSCz galaxies are considered whose clustering properties more
nearly resemble those of the AGNs. This is not surprising given
that most of the brightest galaxies are in fact AGNs, and  the two
samples thus are not truly independent. Regarding the AGNs
sub-samples it can be seen that the clustering of {\sf S1} objects
shows no strong differences to the one of all AGNs; by contrast
the {\sf S2} and {\sf S3} subtypes show a stronger
auto-correlation on the smallest scales, $\theta\alt 3^\circ$.
Note, however, that  the AGN samples, having a smaller number of
objects, have  in general also larger error bars; in this case,
since Poisson statistics makes the errors on $w(\theta)$ decrease
for increasing $\theta$, intermediate scales $\theta\sim
10$--$30^\circ$ might be optimal to distinguish between different
sources. This is especially true for UHECRs when the statistics is
very limited and/or the smearing at the smallest scale by magnetic
fields are important.

The above results are in general quite in good agreement with
other more detailed studies of the AGNs bias properties. In
particular, the clustering properties of AGNs have been studied
extensively for example in Kaufmann {\it et al.}
\cite{Kauffmann:2003dw} using the SDSS catalogue. Their findings
are that AGNs are far more common in massive galaxies and that the
AGN correlation function resembles that of massive early-type
galaxies, which is similar to what we find for the low-redshift
VCV sample.

\section{Comparison with the data and forecast for Auger}\label{compdata}
We turn now to study the clustering of the various source samples
considered at rather small scales, comparing them with the existing
observations. In particular, we shall focus on one of the most
remarkable findings reported by the PAO, namely that of ``8 doublets
within 7 degrees out of the 19 highest energy events'' \cite{ICRC}.

\begin{figure*}[!htbp]
%
\begin{tabular}{cc}
\epsfig{file=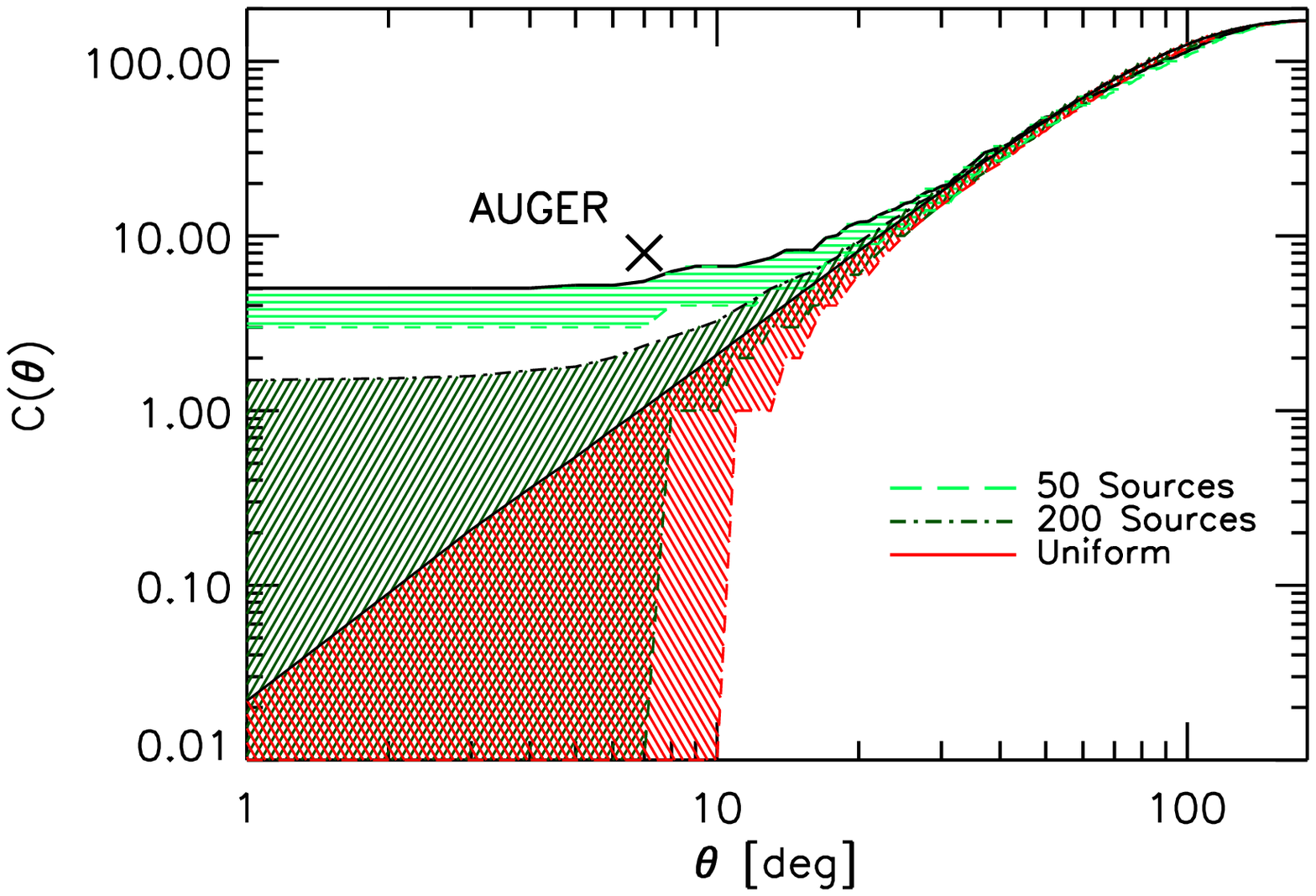,width=1.0\columnwidth} &
\epsfig{file=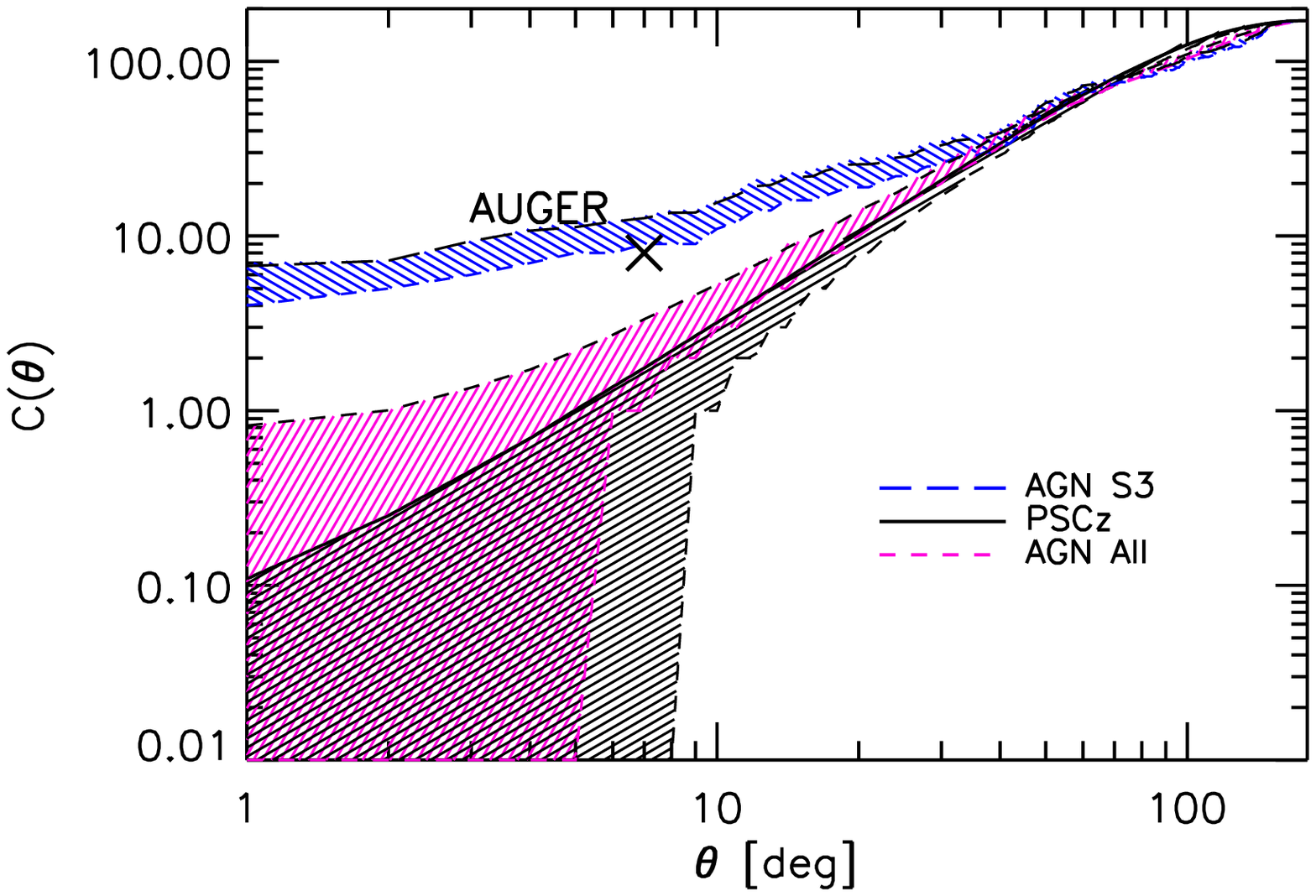,width=1.0\columnwidth} \\
\epsfig{file=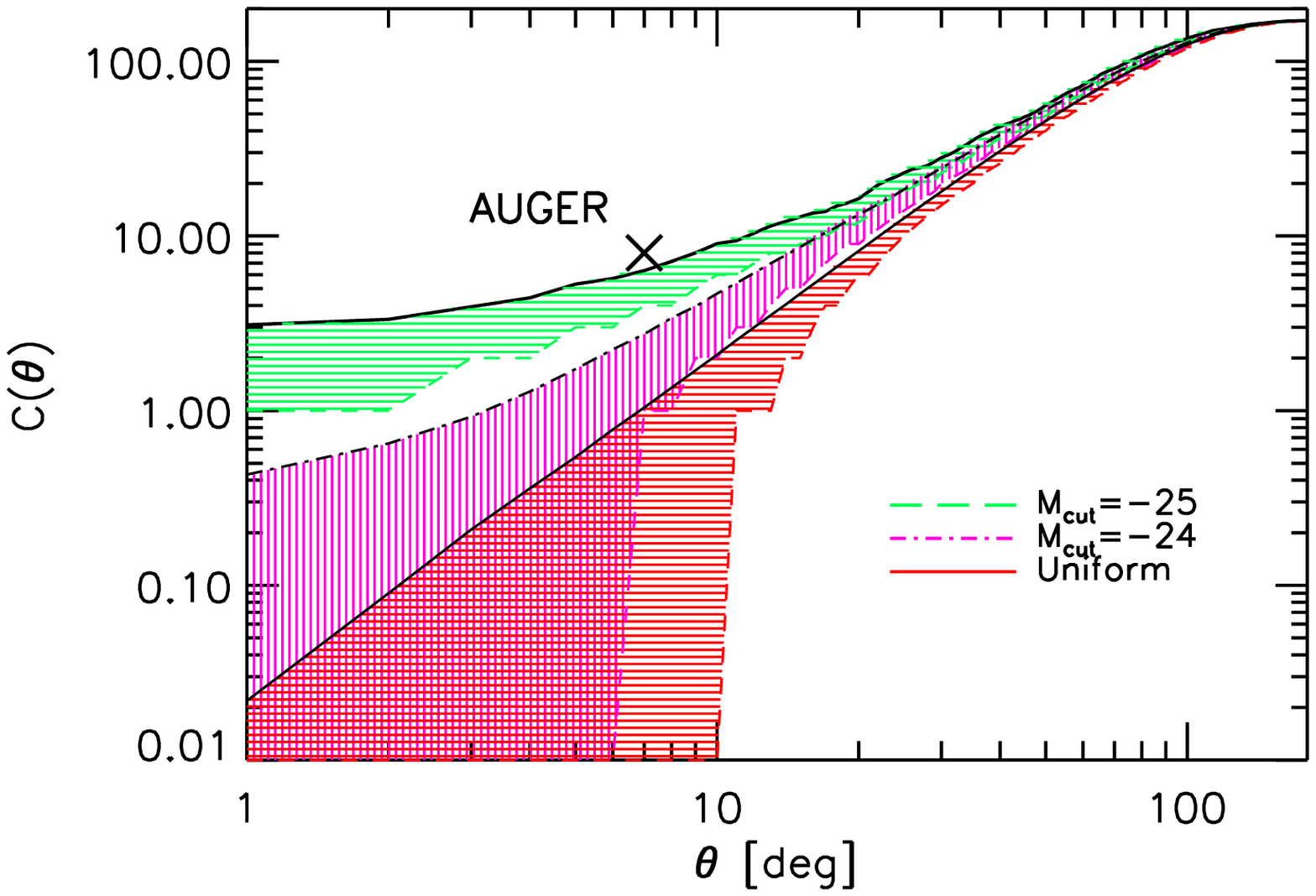,width=1.0\columnwidth} &
\epsfig{file=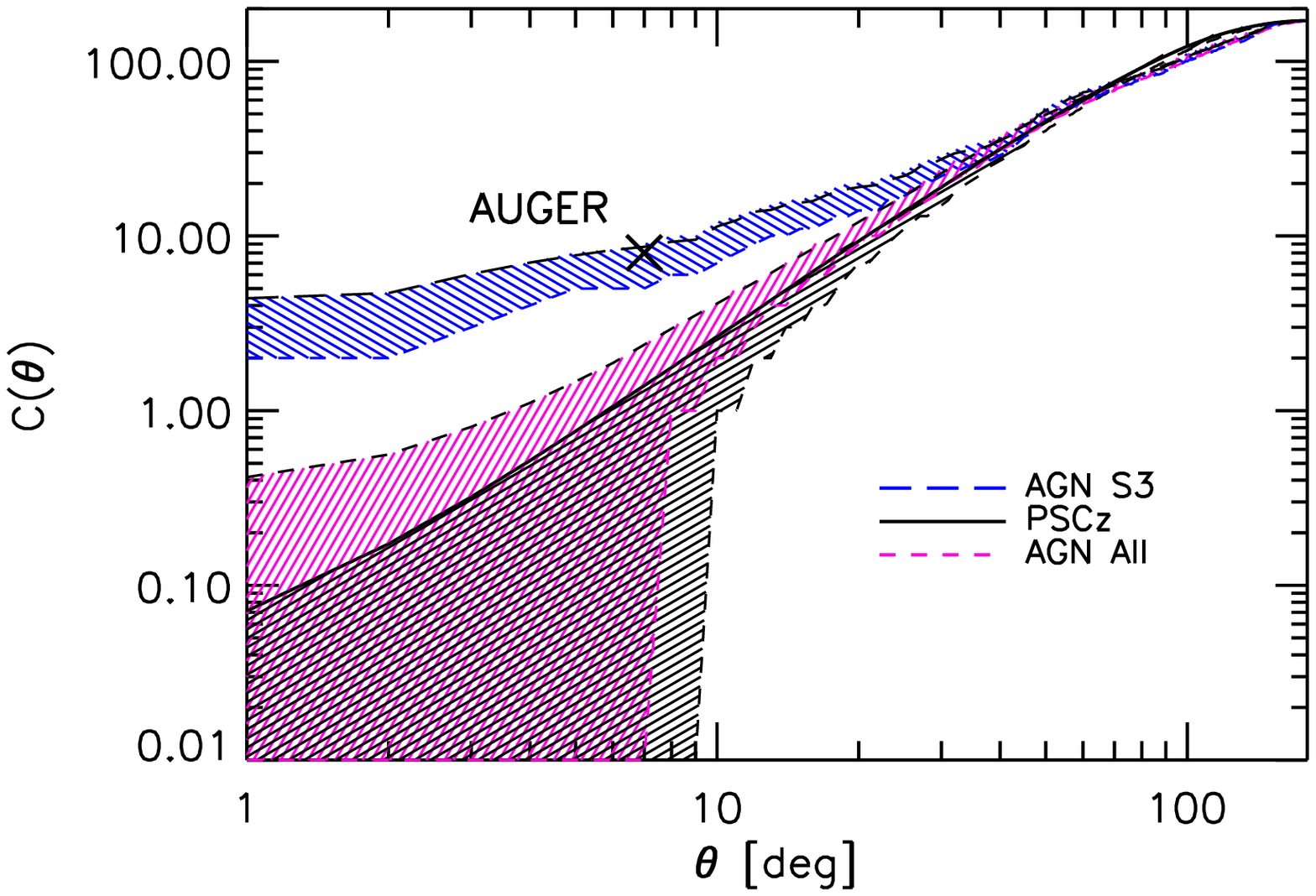,width=1.0\columnwidth}
\end{tabular}
%
\vspace{-1pc} \caption{The average $\C(\theta)$ and its 1 $\sigma$
variation for the case of 19 events and: finite number of sources
isotropically distributed compared with the continuous limit
(top-left panel); Galaxies with different cuts in magnitude
(bottom-left panel); AGN sub-classes compared with Galaxy
distribution (top-right panel); all the previous cases assume a
cut in the window function at $\Ecut=8\times 10^{19}\,$eV. The
bottom-right panel is the same of the third one, but for
$\Ecut=5.75\times 10^{19}\,$eV. Notice that the error regions are
highly asymmetric for $\theta \leq 40^\circ$ (see Figs.
\ref{fig:5} -\ref{fig:7}) and the upper $1 \sigma$ error regions
shown in the plots almost coincide with the mean.}\label{fig:4}
\end{figure*}

Because of the limited UHECR statistics available, and to have a
direct comparison with the Auger findings, in this section we
shall use as main observable $\C(\theta)$ a slightly modified
version of the function $w(\theta)$ of the previous section,
defined as:
\begin{equation}
    \C(\theta)=\left\langle \sum_{i=2}^N\sum_{j=1}^{i-1}\Theta(\theta-\theta_{ij}),\right\rangle \,,
\end{equation}
i.e. the cumulative number of pairs within the angular distance
$\theta$, where $\Theta$ is the step function (with
$\Theta(0)=1$), $N$ the number of CRs considered, and
$\theta_{ij}$ is the angular distance between events $i$ and $j$.
Although $\C(\theta)$ introduces further correlations between
different angular bins, the use of cumulative countings instead of
differential ones has the great advantage of significantly
reducing the dependence from unknown magnetic field deflections, a
crucial point for UHECRs astronomy. The ensemble average is
performed over a large number $M=10^5$ of Monte Carlo sets. The
events are extracted randomly from the catalogue under
consideration and we take into account the PAO exposure as
described in Ref.~\cite{Sommers:2000us} assuming as characteristic
parameters for the PAO $\zeta_{\rm max}=60^\circ$  as maximal
zenith angle for a CR event and $\delta_{\rm PAO}=-35^\circ$ for
the PAO latitude location. The selection effects of the catalogue
and the attenuation due to CR propagation are included assigning
proper weights for the sources, which in turn are used as emission
probabilities in the simulation. Here a hypothesis on the nature
of the particles and on the overall energy scale enters. To
illustrate this point, in Table \ref{tab:1} we report the distance
$D_{1/2}$ from which 50\% of the UHECR flux comes, for different
assumptions about energy and chemical composition, assuming
uniformly distributed sources and rectilinear propagation (see
e.g. \cite{Harari:2006uy}). Note that the injection spectral index
has only a minor effect on $D_{1/2}$ in the energy range
considered.

\begin{table}[b]
\begin{tabular}{|c|c|c|c|}
\hline Species & $\Ecut/10^{19}\,$eV & $D_{1/2}$/Mpc & $z_{1/2}$
 \\ \hline
 \hline $p$ & 5.0 & 160 & 0.037\\
 \hline $p$ & 6.0 & 100 & 0.023\\
 \hline $p$ & 8.0 & 40 & 0.009\\
 \hline ${}^{28}$Si & 6.0 & 30 & 0.007 \\
 \hline ${}^{56}$Fe & 6.0 & 80 & 0.019\\
 \hline ${}^{56}$Fe & 8.0 & 45 &0.011\\
\hline
\end{tabular}
\caption{\label{tab:1}The distance $D_{1/2}$ (or equivalently
redshift $z_{1/2}$) within which 50\% of the UHECR flux above
$\Ecut$ comes for different assumptions on energy and chemical
composition, assuming isotropic and uniform sources and rectilinear
propagation. Adapted from plots in \cite{Harari:2006uy}.}
\end{table}

\begin{figure*}[!th]
\begin{center}
\begin{tabular}{cc}
\epsfig{file=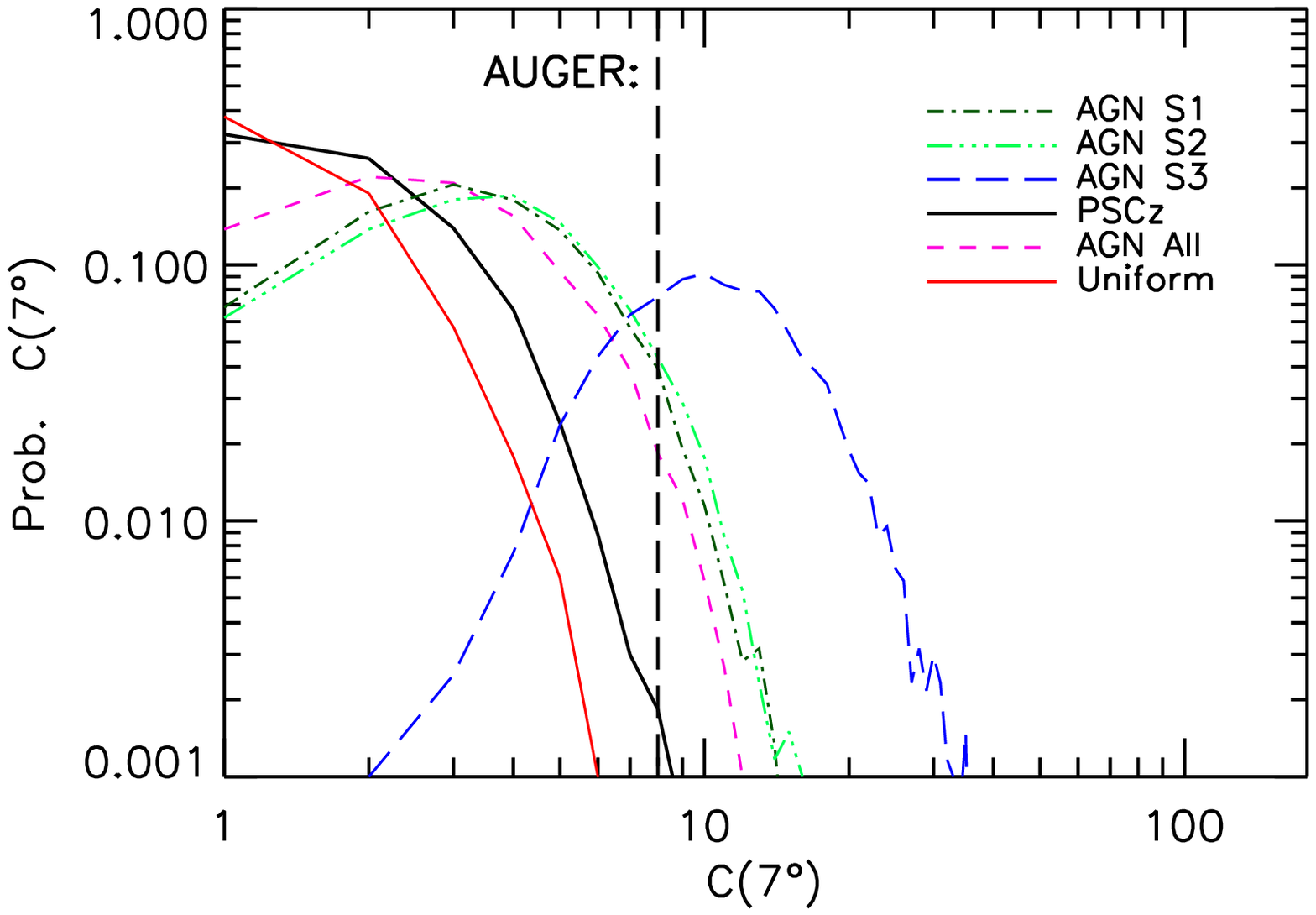,width=0.5\textwidth} &
\epsfig{file=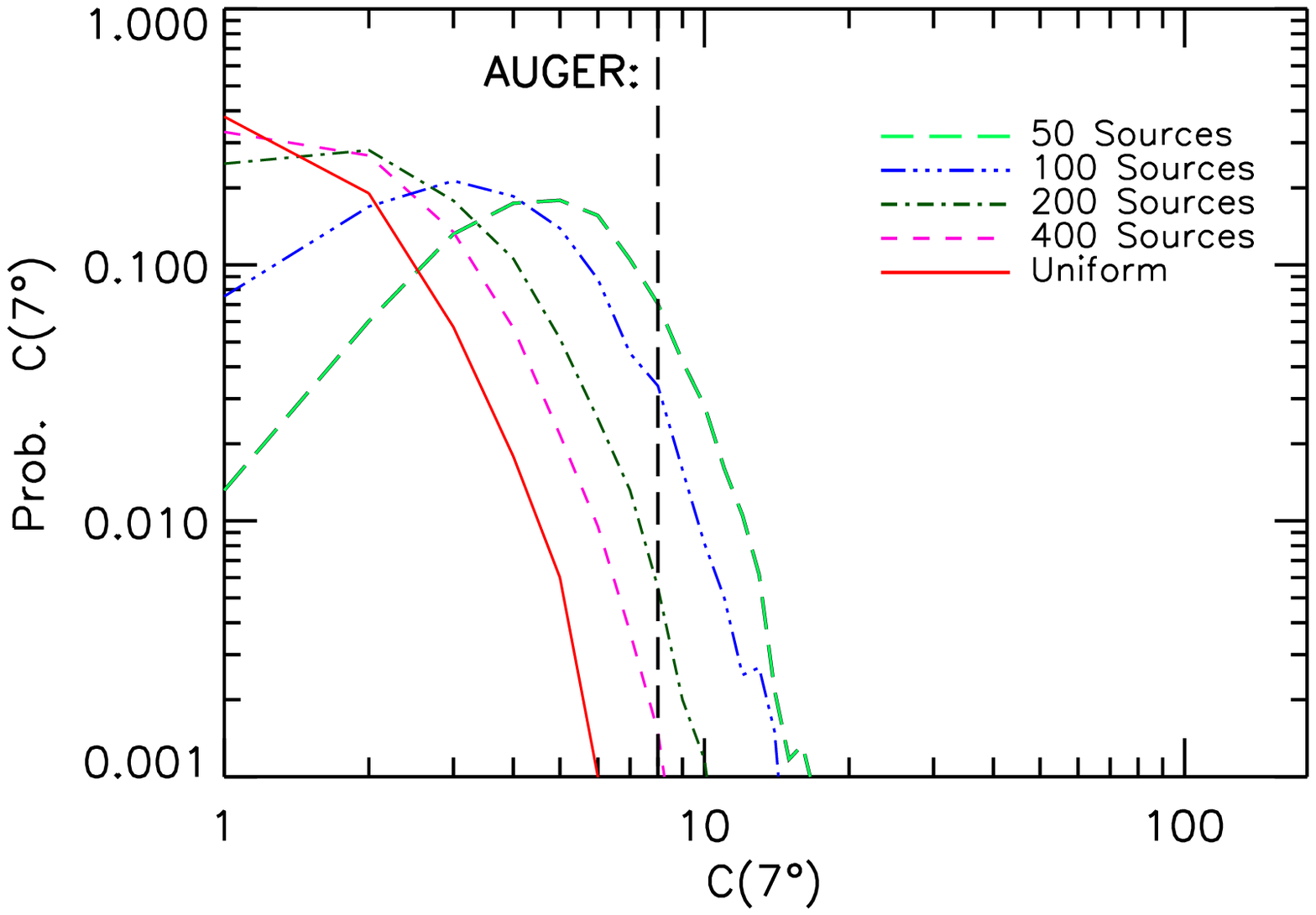,width=0.5\textwidth}
\end{tabular}
\end{center}
\vspace{-1pc} \caption{Probability distribution of $\C(7^\circ)$,
for the case of 19 events, an energy cut of
$\Ecut=8\times10^{19}\,$eV and the case of different astrophysical
models considered (left panel) and a finite number of uniformly
distributed sources (right panel).}\label{fig:5}
\end{figure*}

\begin{figure*}[!htbp]
\begin{center}
\begin{tabular}{cc}
\epsfig{file=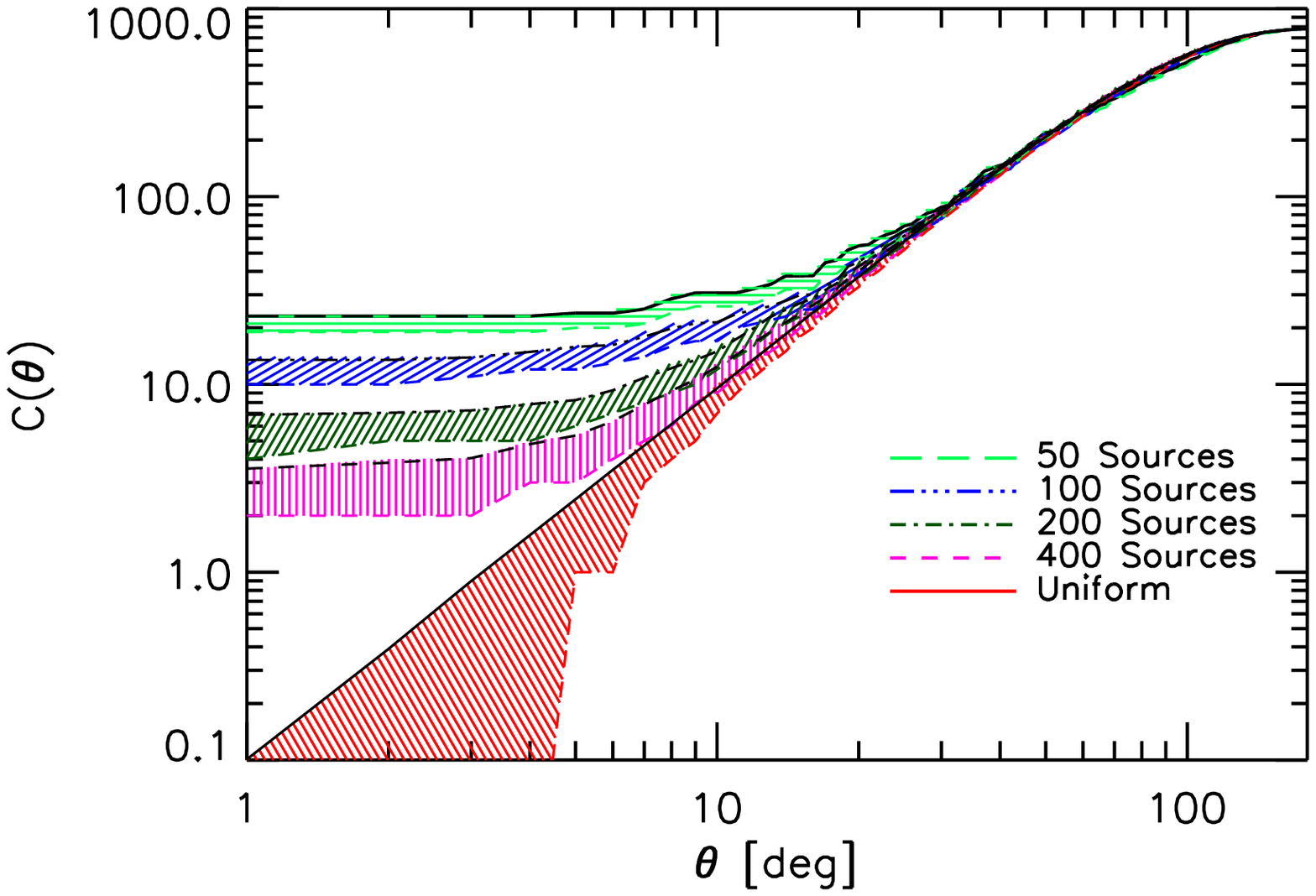,width=0.5\textwidth} &
\epsfig{file=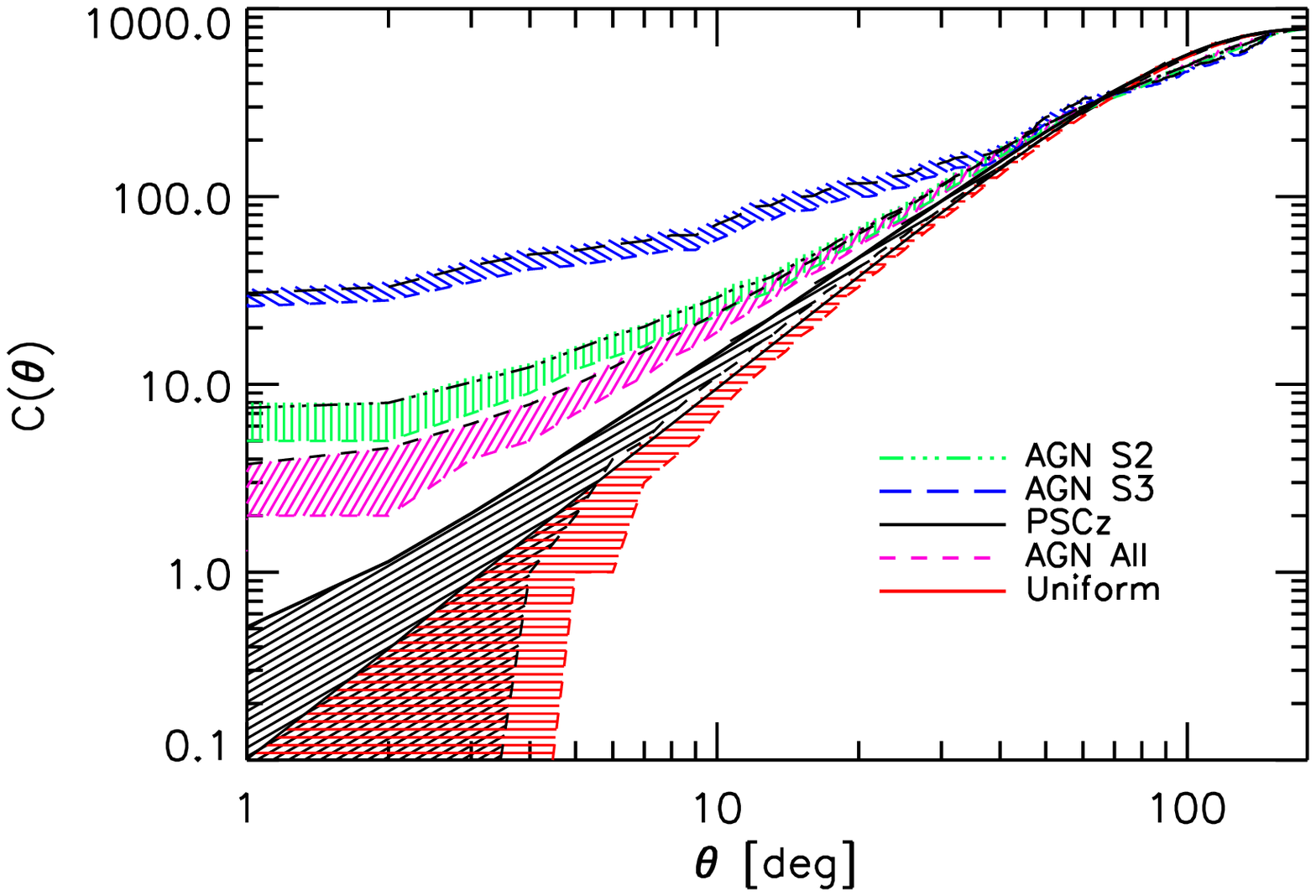,width=0.5\textwidth}
\end{tabular}
\end{center}
\vspace{-1pc} \caption{As in top panels of Fig. \ref{fig:4}, but
for a statistics of 40 events.}\label{fig:6}
\end{figure*}

In the following we assume proton primaries and use the propagation
window function $W(z,\Ecut)$ as calculated in Ref.~\cite{nI}. Given
the importance of the clustering signal observed by the PAO, we shall
focus mostly
on the case of $N=19$ events. In order to study the sensitivity to
the assumed energy scale, we shall consider two cases: (i) the
preliminary calibration of the energy scale presented by the PAO is correct,
and thus the 19 highest energy events have energies above
$\Ecut=5.75\times 10^{19}\,$eV; (ii) UHE air-shower experiments are
affected by an overall uncertainty in their energy calibration, whose
normalization might be obtained however by requiring that they
reproduce correctly spectral features of a model, in our case the
dip model~\cite{dip}. The correction factor found with this method
is 1.4 for the PAO energy scale, in which case the highest energy
events have energies above $\Ecut\simeq 8\times 10^{19}\,$eV.

For the galaxy samples we use the PSCz selection function, while
for the AGN samples we adopt the approximation
$\psi(z)=z^\alpha/z^2$ with $\alpha\simeq 0.4$ as the best-fit
slope to the redshift distribution of the samples (see
Fig.~\ref{fig:2}). Although rather crude, this simple approach can
be justified by the fact that, especially at relatively small
angular scales, the clustering properties are only slightly
affected by the exact choice of the selection and propagation
weights. As a check, we verified that even the extreme choice of
neglecting all selection and propagation effects does not change
appreciably the expected mean and distribution of the number of
pairs. This approximation breaks down when the UHECR horizon
becomes much larger than the distance up to which a catalogue is
complete, a point we shall come back to later.

Our results for the average $\C(\theta)$ and their 1~$\sigma$
variations for the case of 19~events are shown in
Fig.~\ref{fig:4}. From top-left to bottom-right, we present the
case of a finite number of sources uniformly distributed together
with the continuous limit;  AGN sub-classes compared with galaxy
distributions and an isotropic sky; and galaxies with different
cuts in magnitude. While the first three panels assume
$\Ecut=8\times 10^{19}\,$eV, the bottom right panel is the same as
the second one, but for $\Ecut=5.75\times 10^{19}\,$eV. We note
that the strong clustering observed by the PAO is quite
exceptional, and both a uniform random distribution (corresponding
to the limit of an infinite number of sources) and the galaxy
distribution predict in general a too small number of pairs within
7$^\circ$. Active galactic nuclei and in particular their
sub-samples are much more likely to produce the degree of
clustering observed by the PAO. The same happens for the
sub-samples of bright galaxies where the brightest galaxies (and
thus the set with the smallest number density $n_s$) provide the
best match to the expected clustering. In case of a lower energy
scale, the horizon is larger, the sky more isotropic, and
especially the LSS and the isotropic sky hypothesis have even more
trouble in explaining the observations.  Notice however that the
{\sf S3} sample, which seems the AGN sub-sample most consistent
with the high number of pairs, may suffer from a strong selection
bias in the VCV catalogue: LINERs are comparatively weak AGNs,
which are preferentially detected at low $z$ (see Fig.~2). An
additional problem is that most LINERs are outside the field of
view of the PAO, and since their total number is much smaller than
the one of {\sf S1} and {\sf S2} AGNs, cosmic variance plays a
significant role (as for any other sample made of a small number
of objects). Although not manifest from the plots of
Fig.~\ref{fig:4}, another caveat is that, apart from the isotropic
case with an infinite number of sources, virtually all models are
consistent with the observations at the 3~$\sigma$ level. The
exact confidence levels are illustrated in Fig.~\ref{fig:5}, where
the full distribution within $7^\circ$ for various model are
compared to the Auger result. Also, we had to concentrate on the
largest fluctuation in the PAO data-set, since this is the only
presently available information. At different angles, we must
expect less significant clustering. That said, it is interesting
that, as shown in Fig.~\ref{fig:6}, with a statistics doubled with
respect to the one analyzed in \cite{ICRC} the errors should
become sufficiently small to rule out most cases.

Clearly the discrimination power between different source models
would be greatly improved, if the expected functions $\C(\theta)$
were compared to UHECR data not only at a single angular scale but
on a range of values. Already a comparison of the correlations at
a second angle may be enough to distinguish among different cases.
It is likely that a global comparison (based e.g.\ on a $\chi^2$
method or a Kolmogorov-Smirnov test) of the correlation functions
would provide a powerful diagnostic tool. This is one of the main
results of our work and deserves a specific example. For the
present purposes, it is sufficient to illustrate this point by
studying the distribution of the expected number of pairs within
7$^\circ$ (to stick to the most notable finding of the PAO) and
30$^\circ$ (a typical intermediate scale) for several models. In
particular, in table~\ref{tab:2}  we report the average values
$\C(7^\circ)$ and $\C(30^\circ)$ and the 2 $\sigma$ lower and
upper limits---denoted by $\C_{-}$ and $\C_{+}$ respectively---for
the expected number of pairs in different models and $N=19$ data.
In table~\ref{tab:3} and table~\ref{tab:4} we report analogous
quantites for  $N=40$ and $N=60$, respectively. We consider three
models: (a) 100 uniformly distributed sources, which mimicks GRBs;
(b) the {\sf S2} subclass of AGNs; (c) the galaxies brighter than
$M_{\rm cut}=-24.5$. These have been chosen to be basically
consistent with the $7^\circ$ Auger data. Noticeably, even with
the 19 events the models (a) and (b) show significant differences
at $30^\circ$, a difference that become quite large and easily
testable with a modest improvement in statistics to 40 events. The
latter two models are instead almost degenerate from the point of
view of clustering properties, which does not come as a surprise
since the two samples have a similar number of objects and most of
them fall in both subsamples (i.e., they are not independent).
Note further that the distributions are generally quite
non-Gaussian with a prominent tail toward an higher number of
pairs.

\begin{table}[!thb]
\begin{tabular}{|c||c|c|c||c|c|c|}
\hline Model & $\C_{-}(7^\circ)$ & $\C(7^\circ)$ &
$\C_{+}(7^\circ)$& $\C_{-}(30^\circ)$ & $\C(30^\circ)$ &
$\C_{+}(30^\circ)$
 \\ \hline
 \hline 100 GRBs & 0 & 3 & 8 & 10 & 17 & 19\\
 \hline \textsf{S2} AGNs & 0 & 3 & 9 & 12 & 21 & 35\\
 \hline $M_{-24.5}$ Gals & 0 & 3 & 9 & 13 & 25 & 39\\
\hline
\end{tabular}
\caption{Observables related to the distribution of the expected
number of pairs within 7$^\circ$ and 30$^\circ$ and $N=19$ events
(see text for details on the notation). The different models
reported are: 100 GRBs, \textsf{S2} AGNs, Galaxies with $M_{\rm
cut}=-24.5$.\label{tab:2}}
\end{table}

\begin{table}[!thb]
\begin{tabular}{|c||c|c|c||c|c|c|}
\hline Model & $\C_{-}(7^\circ)$ & $\C(7^\circ)$ &
$\C_{+}(7^\circ)$& $\C_{-}(30^\circ)$ & $\C(30^\circ)$ &
$\C_{+}(30^\circ)$
 \\ \hline
 \hline 100 GRBs & 10 & 21 & 32 & 68 & 82 & 96\\
 \hline \textsf{S2} AGNs & 9 & 18 & 31 & 85 & 105 & 151\\
 \hline $M_{-24.5}$ Gals & 6 & 13 & 25 & 81 & 110 & 162\\
\hline
\end{tabular}
\caption{As in Table \ref{tab:2}, but for $N=40$
events.\label{tab:3}}
\end{table}

\begin{table}[!thb]
\begin{tabular}{|c||c|c|c||c|c|c|}
\hline Model & $\C_{-}(7^\circ)$ & $\C(7^\circ)$ &
$\C_{+}(7^\circ)$& $\C_{-}(30^\circ)$ & $\C(30^\circ)$ &
$\C_{+}(30^\circ)$
 \\ \hline
 \hline 100 GRBs & 27 & 40 & 65 & 161 & 194 & 233\\
 \hline \textsf{S2} AGNs & 26 & 44 & 67 & 207 & 261 & 320\\
 \hline $M_{-24.5}$ Gals & 21 & 32 & 52 & 195 & 268 & 330\\
\hline
\end{tabular}
\caption{As in Table \ref{tab:2}, but for $N=60$
events.\label{tab:4}}
\end{table}

As a final comment, we notice that due to the limited information
available we have restricted the study to the analysis of
cumulative number of pairs. However, the above approach can be
easily generalized to higher order statistics, like the cumulative
counting of the proper number of doublets, of triplets, etc.
\cite{Harari:2004py}. A combined use of these tools will likely
provide even a more robust and stringent constrain on the nature
and number of UHECR sources.

\subsection{Repeaters vs. small scale clustering}
An important issue for the future study of UHECR sources is to
disentangle the case where an excess of pairs should be attributed to
multiple events from a single point source from the case where the
excess is produced by the small-scale correlation of two or more
sources, as discussed in the previous sections for AGNs and bright
galaxies.

\begin{figure}[!t]
\begin{center}
\epsfig{file=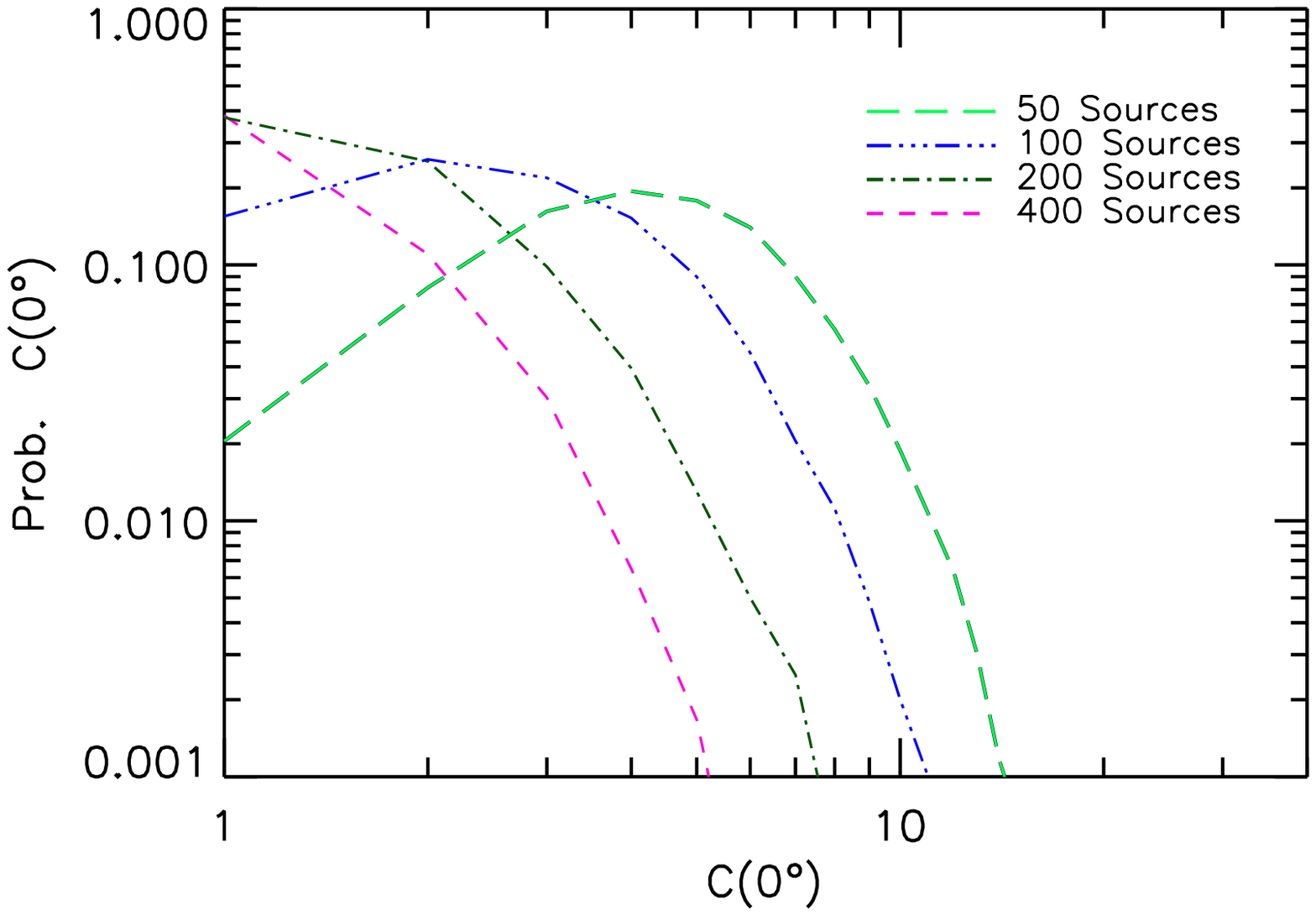,width=0.5\textwidth} \\
\epsfig{file=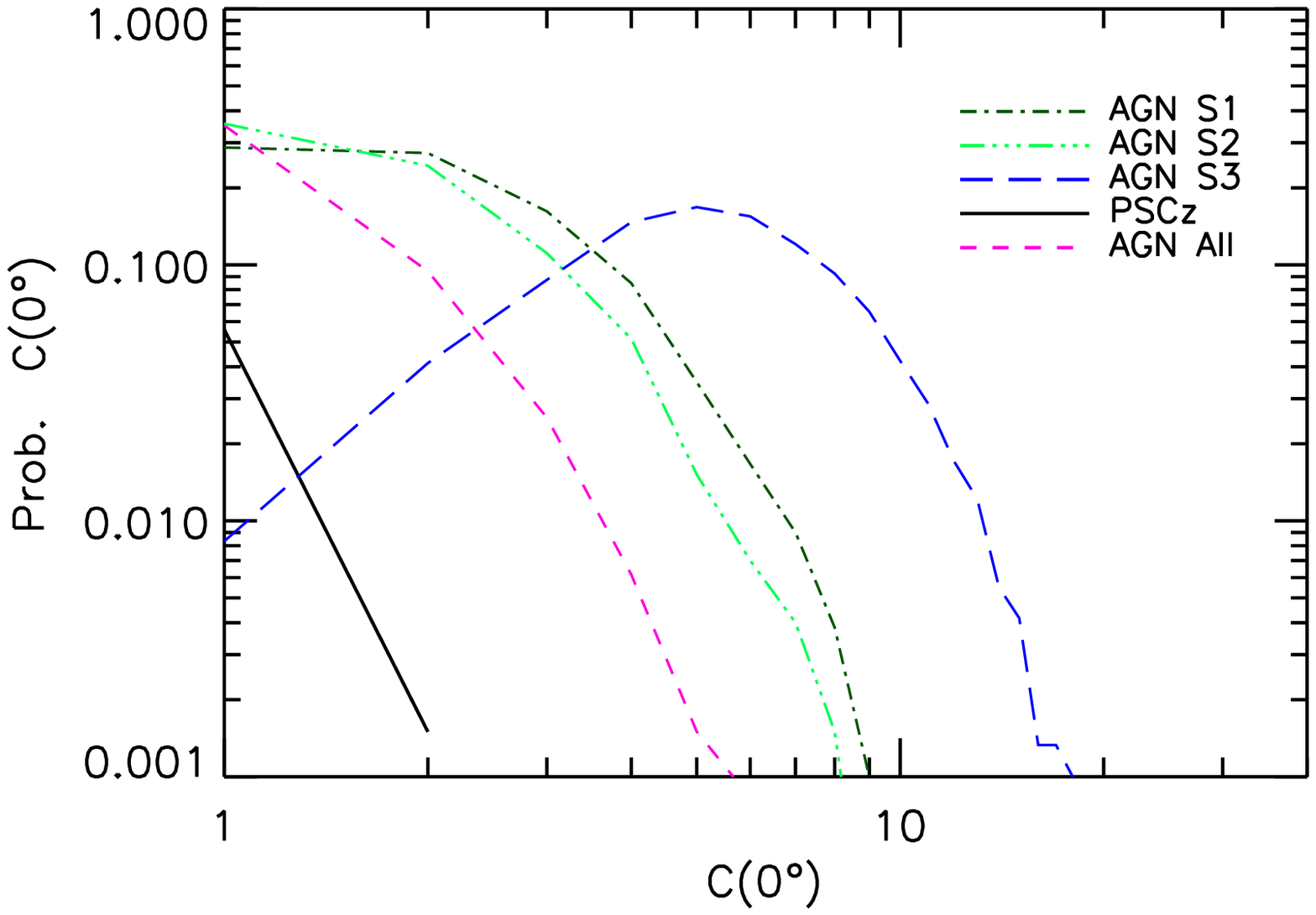,width=0.5\textwidth}
\end{center}
\vspace{-1pc} \caption{Probability distribution of $\C(0^\circ)$,
for the case of 19 events, an energy cut of
$\Ecut=8\times10^{19}\,$eV and a finite number of uniformly
distributed sources (top panel) and the case of different
astrophysical models considered (bottom panel). Notice that
uniformly distributed sources predicts $P <10^{-3}$ for
$C(0^\circ)=1$. The related curve would thus not be visible in the
plots.}\label{fig:7}
\end{figure}

For a large fraction of the models considered above, the predicted
clustering is mostly due to the intrinsic correlations of the
sources rather than being caused by multiple emissions from single
sources\footnote{As a consequence, we note that a strong intrinsic
  small scale clustering within the experimental angular resolution
might prevent an unambiguous identification of the source of a given
event even in absence of magnetic deflections.}.
A formal but useful way to illustrate this point is to look at the
probability distribution of $\C(0^\circ)$, i.e. the number of pairs
for $\theta=0^\circ$ (repeaters). In Fig.~\ref{fig:7} we report this
distribution for the case of 19 events and an energy cut of
$\Ecut=8\times10^{19}\,$eV. We show both the case of a finite number
of uniformly distributed sources and the case of different
astrophysical models considered. In the former case, there should be
less  than about 50~UHECR sources within the horizon in order
to observe a dominant fraction of repeaters as origin for the doublets.

Of course, deflections in magnetic fields and experimental resolution
effects prevent an experimental determination of $\C(0^\circ)$. As
an example, we ask if the strong clustering signal in the PAO
data coming from pairs within 7$^\circ$ is compatible with repeaters.
The following arguments show that,
without additional information, both the hypotheses of repeaters and
of small-scale clustering are consistent with the findings. The
correlation functions $w(\theta)$ of the astronomical catalogues
previously discussed are peaked at small scales, and actually
strongly peaked below one degree for the AGN samples. Yet, this peak
in $w(\theta)$ will be shifted in the excess signal in $\C(\theta)$
to larger scales, first of all because the event numbers scales
approximately as $N\propto\theta^2$. Moreover, we expect the
small-scale signal in the data to be washed away partly because of
the angular resolution of the detector (that in the PAO data-set
mentioned is $\sim 1^\circ$), and mostly because of deflections in
the Galactic magnetic field (GMF) and possibly extragalactic ones.
Even protons of the considered energy are expected to suffer average
deflections in the GMF of order $\sim 3^\circ$~\cite{kst}. Since the
Auger exposure peaks near the Galactic Center region, which in
typical GMF models is associated with larger deflections,
separations of point-like sources of protons up to $7^\circ$ by the
GMF alone cannot be excluded. Other reasons for the large separation
angle are deflections in extragalactic magnetic
fields, or an intermediate-heavy chemical composition of UHECRs that
would lead to increased deflections in the GMF.

These arguments emphasize once more the importance of a global
comparison of the auto-correlation function to perform a robust
diagnostics.

\subsection{Comparison with old experiments}
Given the importance of a global comparison of the auto-correlation
function, one may wonder if the already existing publicly available
data offer additional insight. The only sufficiently large data set
that is publicly available is the one used for the first time in
Ref.~\cite{msc}. It consists of $\sim 100$ events with energies
$E\geq 4\times 10^{19}\:$eV from the HiRes
stereo~\cite{Abbasi:2004ib,Abbasi:2004ib_b}, AGASA~\cite{AG},
Yakutsk~\cite{YK} and SUGAR~\cite{SG} experiments. Here, we have
rescaled the absolute energy scale of each
experiment~\cite{msc,ks03} by requiring that they reproduce
correctly the dip spectral feature~\cite{dip}.

\begin{figure}[!th]
\begin{center}
\epsfig{file=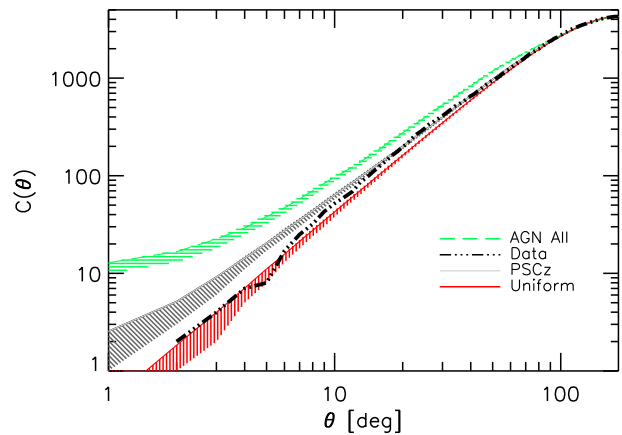,width=0.5\textwidth}
\end{center}
\vspace{-1pc} \caption{Cumulative number of pairs and  $1\,\sigma$
error regions as in Figs.\ref{fig:4},\, \ref{fig:6}  for the set
of pre-Auger CR data compared with  AGNs and PSCz galaxies within
$z=0.02$, and with a uniform expectation.}\label{fig:8}
\end{figure}

Unfortunately, even at energies $E\sim 5\times 10^{19}\:$eV, for
the case of protons and a rectilinear propagation, sources beyond
redshift $z\sim 0.04$ contribute about half of the flux, see
Table~\ref{tab:1}. At those distances the catalogues are known to
be incomplete, and although we correct for the selection function,
we cannot apply the method previously outlined in a reliable,
quantitative way. Nonetheless, in Fig.~\ref{fig:8} we compare for
illustrative purposes the function $\C(\theta)$ computed for this
data-set with the corresponding expectations from uniformly
distributed sources and for AGNs as well as the PSCz Galaxy
Catalogue within $z\alt 0.02$.

It can be seen that the auto-correlation of the data presents an
excess at $10^\circ-30^\circ$ with respect to a uniform
distribution corresponding to the medium scale clustering signal
found in \cite{msc}. Although a quantitative analysis would likely
provide a poor fit, one instead recognizes a qualitative
similarity in the pattern of the data function and that of the
samples following the LSS, i.e. AGNs and galaxies. (Note, however,
that a comparison below $\theta\simeq 10^\circ$ is not very
meaningful, given the poor angular resolution of SUGAR, for
example.) Definitely, more statistics at higher energy and better
quality data are needed from UHECR experiments, while deeper and
more complete large-angle surveys would be welcome from the
astrophysical community.

\section{Discussion and conclusions}\label{conclusion}
We have examined the clustering properties of ordinary galaxies,
GRBs and AGNs with the aim of finding characteristic features
which may shed light on them as possible UHECR sources. Our
auto-correlation studies have shown that---consistently with what
is known from the much larger SDSS galaxy and AGN catalogues, see
e.g.~\cite{Kauffmann:2003dw,Constantin:2006uu,Kewley:2006ft,Best:2006xz}---nearby
AGNs exhibit much stronger small-scale clustering than average
galaxies. The same is true for the brightest galaxies in the PSCz
catalogue that are mainly big ellipticals (plus some starburst
galaxies).  Since many of them do overlap with known AGNs, these
two samples are not truly separate and the similarities in the
small-scale clustering of bright galaxies and AGNs are not
surprising. Unfortunately, both the overdensity overlap of
physically different classes of sources and the pronounced
small-scale clustering of many source candidates does not play in
favor of a clear source identification of UHECR sources e.g.~by
cross-correlation analyses.

We have argued that the auto-correlation function of different
source classes differs considerably on all scales and may be used as
a tool to identify the sources of UHECRs. Since the PAO has not
 yet published sufficient information on their observed events, we were
restricted to perform a more conventional analysis considering
just one bin of the auto-correlation function. At present, the
most likely interpretation of the evidence reported by Auger of
``8 doublets separated by less than 7 degrees in the 19 highest
energy events'' is that the sources of UHECRs are either a
strongly clustered sub-sample of AGNs, or a sparse population of
more or less isotropically distributed sources (e.g.~GRBs),
possibly with pairs of events within 7$^\circ$ coming from the
same objects. From our results, it is however clear that a
comparison on all angular scales would disentangle the two cases.
In principle, once the source population giving rise to UHECRs is
identified, the magnetic field deflection required to smear-out
the original auto-correlation function might be fitted and used
for studies of the (extra-) galactic magnetic field.

With a statistics as low as twice the preliminary sample analyzed by
the Auger collaboration, we expect that a first conclusive
discrimination among source populations should be possible.
Measuring the difference between e.g.\ different subclasses  of AGNs
as sources of cosmic rays appears to be more difficult and requires
more complete catalogues within the near ($z\alt 0.1$) universe and
much larger statistics.

Probably, opening the era of UHECR astronomy will require a
combined advance in many aspects of UHECR physics, from reducing
the uncertainty on the absolute energy scale to robust constraints
on the chemical composition of the primaries. At the same time,
the field would also benefit from advancements in the astrophysics
of magnetic fields, like constraints on the Galactic magnetic
field and refined simulations of extragalactic ones. No doubt,
however, that once born UHECR astronomy will pay-off as an
unprecedented diagnostic tool for the study of the high energy
non-thermal universe, as well as for measuring otherwise
inaccessible extragalactic magnetic fields.

\section*{Acknowledgments}
A.C., S.H. and M.K.\ thank the Max-Planck-Institut f\"ur Physik in
Munich for hospitality and support during the initial phase of this
work. P.S. acknowledges support by the US Department of Energy and
by NASA grant NAG5-10842. This work was supported in part by the
European Union under the ILIAS project, contract
No.~RII3-CT-2004-506222.



\end{document}